\documentclass[%
 reprint,
 amsmath,amssymb,
aps,
prl,
floatfix,
]{revtex4-2}

\usepackage{graphicx}
\usepackage{dcolumn}
\usepackage{bm}
\usepackage{hyperref}
\usepackage{xfp}

\begin{document}
\title{A pure and indistinguishable single-photon source at telecommunication wavelength}

\author{Beatrice Da Lio}
\author{Carlos Faurby}
\author{Xiaoyan Zhou}
\altaffiliation[Currently at: ]{Key Laboratory of Integrated Opto-Electronic Technologies and Devices in Tianjin, School of Precision Instruments and Opto-Electronics Engineering, Tianjin University, Tianjin 300072, China.}
\author{Ming Lai Chan}
\author{Ravitej Uppu}
\altaffiliation[Currently at: ]{Department of Physics \& Astronomy, University of Iowa, Iowa City, IA, USA}
\author{Henri Thyrrestrup}
\affiliation{Center for Hybrid Quantum Networks (Hy-Q), Niels Bohr Institute, University of Copenhagen, Blegdamsvej 17, DK-2100 Copenhagen, Denmark.}

\author{Sven Scholz}
\author{Andreas D. Wieck}
\author{Arne Ludwig}
\affiliation{Lehrstuhl f\"ur Angewandte Festk\"oorperphysik, Ruhr-Universit\"at Bochum,
Universit\"atsstrasse 150, D-44780 Bochum, Germany}

\author{Peter Lodahl}
\author{Leonardo Midolo}
\email{midolo@nbi.ku.dk}
\affiliation{Center for Hybrid Quantum Networks (Hy-Q), Niels Bohr Institute, University of Copenhagen, Blegdamsvej 17, DK-2100 Copenhagen, Denmark.
}

\date{\today}

\begin{abstract}
On-demand single-photon sources emitting pure and indistinguishable photons at the telecommunication wavelength are a critical asset towards the deployment of fiber-based quantum networks. Indeed, single photons may serve as \textit{flying qubits}, allowing communication of quantum information over long distances. Self-assembled InAs quantum dots embedded in GaAs constitute an excellent nearly deterministic source of high quality single photons, but the vast majority of sources operate in the 900-950 nm wavelength range, precluding their adoption in a quantum network. Here, we present a quantum frequency conversion scheme for converting single photons from quantum dots to the telecommunication C band, around 1550 nm, achieving 40.8\% end-to-end efficiency, while maintaining both high purity and a high degree of indistinguishability during conversion with measured values of $g^{(2)}(0)=2.4\%$ and $V^{(\text{corr})}=94.8\%$, respectively.
\end{abstract}

\maketitle

Single photons at telecommunication wavelengths constitute a key resource for quantum communication, enabling transmission of quantum information over long distances across optical fiber networks without loss of coherence. 
A complete quantum network would require the ability to exchange information between matter qubits and photons with high fidelity. This demands a deterministic emitter-photon interface capable of realizing photon generation and entanglement in a scalable manner \cite{uppu2021quantum}. In combination with mature foundry-based silicon photonic integrated circuit technology, the realization of a scalable quantum photonic technology is within reach. 
A solid-state quantum emitter deterministically coupled to a cavity or waveguide is an attractive source of on-demand single photons. In particular, self-assembled InAs quantum dots (QDs) in GaAs offer a mature technology for photon generation and a platform for interfacing electron spin and photons in which the decoherence and noise processes have been identified and reduced \cite{uppu2020scalable,appel2021coherent,pedersen2021demand,wang2016near,tomm2021bright}.
Despite the excellent performance, a major drawback of InAs-based QD emitters is that the emission wavelength is typically in the range of 900-950 nm. However, standard single mode fibers experience the lowest attenuation coefficient of 0.16 - 0.2 dB/km at around 1550 nm, while the attenuation in the 900-950 nm range is severely limited by Rayleigh scattering to about 1 - 1.5 dB/km \cite{miya1979ultimate}. Hence, the QDs typical emission wavelength reduce their applications to communication over only short distances and impedes interfacing to advanced quantum photonic integrated circuits on the silicon-on-insulator platform. 

Controlling the epitaxial growth of self-assembled QDs does enable shifting the wavelength of the emitters to longer wavelengths, i.e. to the telecommunications O-band (1300 nm) and C-band (1550 nm), by providing a deeper electron-hole confinement. However, InAs QDs in pseudomorphically strained InGaAs quantum wells \cite{seravalli2007quantum} or grown by metalorganic vapour-phase epitaxy \cite{anderson2020quantum}, typically suffer from high dislocation densities or large background impurity concentration, resulting in higher charge noise, blinking, or low internal quantum efficiency from non-radiative decay channels. In fact, to date, several demonstrations of single-photon emission at telecommunication wavelengths exist \cite{takemoto2004non,miyazawa2005single,ward2005demand,takemoto2007optical,takemoto2008telecom,benyoucef2013telecom,paul2015metal,miyazawa2016single,al2016resonance,paul2017single,muller2018quantum,kolatschek2021bright}, but there are only few reports on the photon indistinguishability \cite{kim2016two,nawrath2019coherence,srocka2020deterministically} where temporal post-selection was implemented at the cost of reduced source efficiency. These works are important stepping stones towards the realization of a quantum network and indeed prepare-and-measure quantum key distribution protocols have already been implemented with such sources, as they do not require indistinguishable photons \cite{intallura2007quantum,intallura2009quantum,takemoto2010transmission,takemoto2015quantum,kupko2021evaluating}. However, the full potential of the technology, e.g., in device-independent quantum key distribution or one-way quantum repeaters, is only realized with high-efficiency on-demand sources of indistinguishable photons in the telecom range.

\begin{figure*}
\centering
\includegraphics[width=0.95\textwidth]{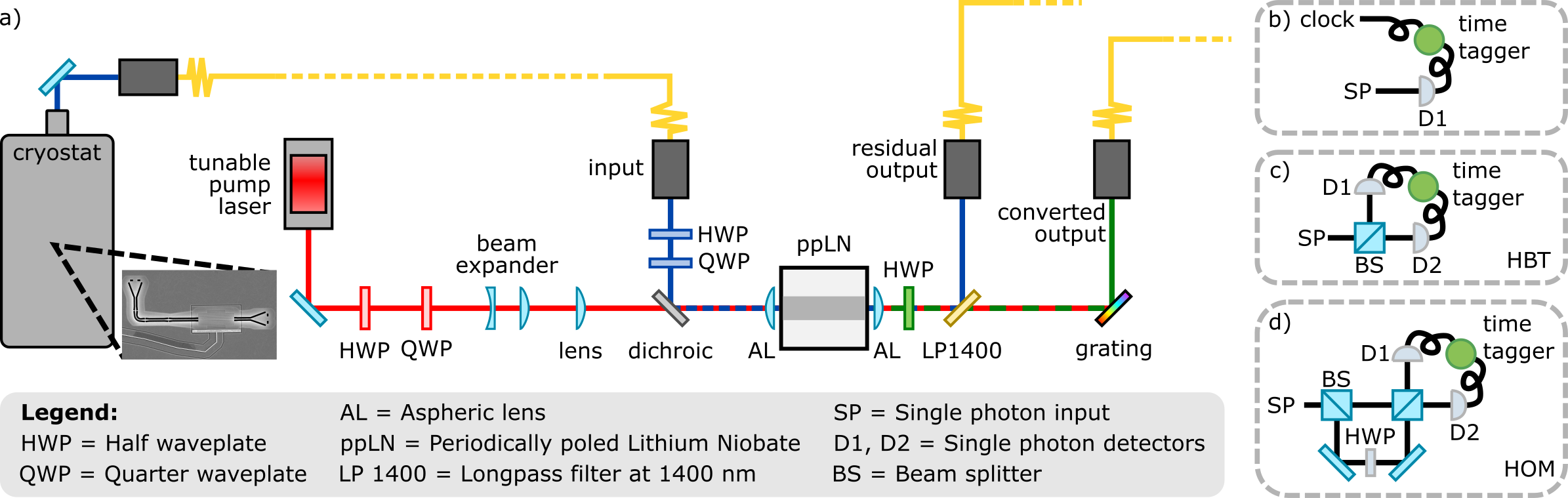}
\caption{\textbf{Schematic of the experimental setup used for frequency conversion and measurements. a) Frequency conversion setup.} Single photons at 940 nm are emitted by a quantum dot in the photonic crystal waveguide on-chip in a cryostat at 1.6 K or 4 K (see main text) and routed to the frequency conversion setup via a single mode fiber. Single photons (blue path) are mixed with a dichroic mirror with light from a tunable continuous wave laser at around 2400 nm (red path), and both are coupled to the ppLN waveguide via an aspheric lens (AL), after accurate mode matching of both polarization (using HWP and QWP) and mode profile (using a beam expander and an extra lens). At the output, the converted single photons (at 1550 nm, green path) are collimated by another AL, and all unwanted residual wavelengths are filtered away with a longpass filter at 1400 nm and a diffraction grating, before the coupling to a single mode fiber. \textbf{b) Photon counting and lifetime measurement setup.} Single photons (either before or after conversion) are sent to a single-photon detector and counts are collected by a time tagger unit. For the lifetime measurement, a clock from the excitation laser is used for synchronization. \textbf{c) Purity measurement setup.} Single photons are directed to a Hanbury Brown-Twiss (HBT) setup composed of a beam splitter with a single-photon detector at each output, and a time tagger unit records count coincidences between them. \textbf{d) Indistinguishability measurement setup.} Single photons are directed to a Hong-Ou Mandel (HOM) setup composed of a Mach-Zehnder interferometer with delay matching the repetition rate of the excitation laser and a HWP to cross-polarize photons. Two single-photon detectors at the output and a time tagger unit measure coincidences.}
\label{fig:setup}
\end{figure*}

An alternative route to reach this goal exploits quantum frequency conversion (QFC) of the emitted single photons to the C-band, thereby  bridging the gap between the mature InAs emitter technology and quantum communications applications. Previous works have shown efficient QFC to either the O-band \cite{zaske2012visible} or the C-band \cite{pelc2012downconversion,Kambs:16} while maintaining the purity of the single-photon source. Recently, Morrison et al. \cite{morrison2021bright} also analysed the indistinguishability after QFC and reported two-photon interference of up to $(67 \pm 2)\%$. 

In this work, we report stable and efficient QFC of single photons from QDs to the telecommunication C-band, with a total $(40.8 \pm 0.8)\%$ system efficiency and simultaneously low multi-photon probability of $(2.4 \pm 0.2)\%$ and high degree of indistinguishability $(94.8 \pm 1.6)\%$ of the converted photons, without implementing any post selection. The conversion method uses difference-frequency generation in periodically-poled Lithium Niobate (ppLN) waveguides, approaching near-unity internal conversion efficiency, which is only limited by fabrication imperfections (e.g. non uniform waveguide cross-section), mode matching between the three waves and scattering. Our QFC scheme enables implementing single-photon based protocols for quantum key distribution over long distances and combining high-quality on-demand single-photon sources with photonic integrated circuits based on silicon photonics.


The frequency conversion process implemented in this work is based on difference frequency generation (DFG). DFG is a second-order non-linear process in which photons from two input wavelengths, $\lambda_1$ and $\lambda_2$ are combined to produce output photons at a third wavelength $\lambda_3=1/(1/\lambda_1-1/\lambda_2)$, i.e. at the difference of the input frequencies. Since the InAs/GaAs QDs emit photons in the range of 900 to 950 nm ($\lambda_1$), and the goal is to convert them to the telecommunication C-band around 1550 nm ($\lambda_3$), a pump laser source at around 2100-2500 nm ($\lambda_2$) is required. In this work, we used a 48 mm long ppLN waveguide with 26 $\mu$m poling period.
A schematic representation of the setup used for DFG is shown in Fig.~\ref{fig:setup}a). The setup consists of one input fiber for photons at 940 nm (subsequently collimated to a free-space beam), free-space optics to optimally couple to the desired ppLN waveguide,  and an output fiber where converted telecommunication photons are coupled into. We define the system, or end-to-end, efficiency as measured from fiber to fiber. At the input of the ppLN waveguide, single photons from the QD are mixed with light from a high-power continuous-wave pump laser. At the output, the converted 1550 nm wavelength photons are spectrally filtered and directed to the output fiber. Spectral filtering is crucial to achieve a low-noise process, as required to reach high purity and indistinguishability. Here, we use a longpass filter to remove residual input photons and second harmonic generation from the pump and a blazed grating acts as a narrow bandpass filter to extinguish the pump from the converted output. Further details on the experimental setup are reported in the supplementary materials.

\begin{figure}
\centering
\includegraphics[width=0.4\textwidth]{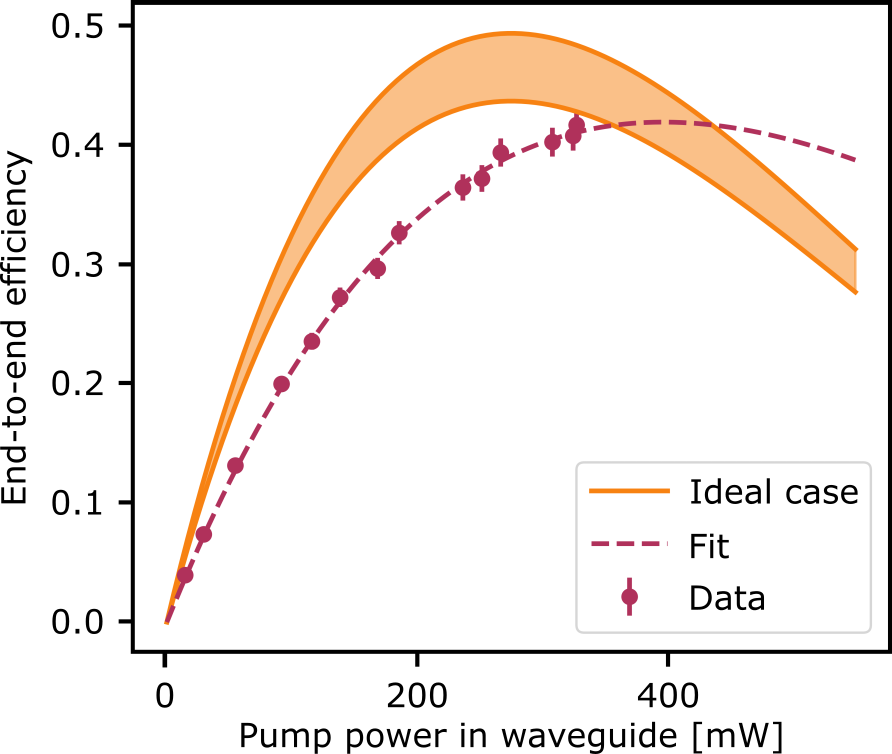}
\caption{\textbf{Pump power dependence of the frequency conversion setup.} External efficiency measured when converting light from a tunable continuous wave laser for a classical characterization of the frequency conversion setup. Data are shown with purple points, the fit with a purple dashed line and the upper and lower bounds for the ideal case with orange lines.}
\label{fig:satPower}
\end{figure}

The system efficiency is characterized with a classical light beam at the input. We use a continuously tunable laser emitting continuous-wave light in the required wavelength range. Varying the pump power coupled to the waveguide, we measured the external conversion efficiency as: $\eta_{ext}=(P_{out} \lambda_3)/(P_{in} \lambda_1)$, where $P_{in}$ and $P_{out}$ are the input and output power measured in fiber at $\lambda_1=940$ and $\lambda_3=1550$ nm respectively. The results are shown in Fig.~\ref{fig:satPower}. 
The highest observed end-to-end conversion efficiency is $(41.7 \pm 0.3)$\%, for 327 mW of pump power in the waveguide, which is close to the maximum efficiency expected from theoretical analysis (see supplementary materials).
Fig.~\ref{fig:satPower} also shows the lower and upper bounds for the ideal case scenario, which are estimated assuming 100\% internal conversion efficiency and including the measured optical loss of the setup (with the uncertainty coming from the in- and out-coupling scattering losses and from the propagation loss of the 1550 nm photons in the waveguide). 
The measured values and the fit show that the maximum conversion efficiency is lower than in the ideal scenario and it is reached at higher pump power. The lower efficiency is explained by a non-unity internal conversion efficiency, which is mainly due to an imperfect coupling of the single photons to the fundamental mode of the waveguide, which supports several modes at 940 nm. We suspect that the requirement for higher pump power to reach saturation can be explained by assuming a varying waveguide cross-section across its length due to fabrication imperfections.

We estimate the internal conversion efficiency using two methods, with each method providing either the upper or the lower bounds. In the first method, we measure the input and output converted power right before and after the aspheric lenses used for in- and out-coupling and then factor out the aspheric lenses insertion loss (6.5\% at both 940 and 1550 nm) and the simulated coupling efficiency (96\%). As we neglect the potential scattering losses occurring at the input and output of the waveguide (e.g. arising from surface roughness), we are estimating a lower bound for the actual internal conversion efficiency $\eta_{\text{int}}^{\text{L}}$. In the second method, we compare the 940 nm laser transmission with pump laser on and off to estimate the depletion due to DFG. As we negelct the optical loss at 1550 nm, we are estimating the efficiency with an upper bound $\eta_{\text{int}}^{\text{U}}$. We measured $\eta_{\text{int}}^{\text{L}}=(86 \pm 0.3)\%$ and $\eta_{\text{int}}^{\text{U}}=(95 \pm 1)$\%.


The QFC experiment is performed with a QD source in a 1.6 K closed-cycle cryostat (see reference \cite{uppu2020scalable} for a description of the source) emitting at 942 nm. The end-to-end efficiency of the QFC is obtained by comparing the single-photon rate at 1550 nm at the output of the setup with the single-photon rate at 930 nm at the input of the setup. Under $\pi$-pulse resonant excitation (with 73 MHz repetition rate and 20 ps pulse duration) we measure an in-fiber rate of N$_{\text{in}}$=$(2.21 \pm 0.02)$ Mcount/s at the fiber input of the frequency conversion setup. The achieved in-fiber count rate at the output of our setup is N$_{\text{out}}$=$(905 \pm 15)$ kcount/s. Both N$_{\text{in}}$ and N$_{\text{out}}$ are obtained measuring the photon rate with SNSPDs
and correcting for the setup-to-detector fiber transmissions and detector efficiencies (see supplementary materials).
This leads to an overall external conversion efficiency of $\eta_{\text{ext}}$=N$_{\text{out}}$/N$_{\text{in}}$=$(40.8 \pm 0.8)$\%. The achieved $\eta_{\text{ext}}$ is fully consistent with the value obtained for classical input light, as discussed previously, meaning that the bandwidth of the QD (measured to be around 500 MHz) is sufficiently narrow to be covered by the frequency conversion process bandwidth, whose full width half maximum is measured to be 115 GHz.

\begin{figure}
\centering
\includegraphics[width=0.45\textwidth]{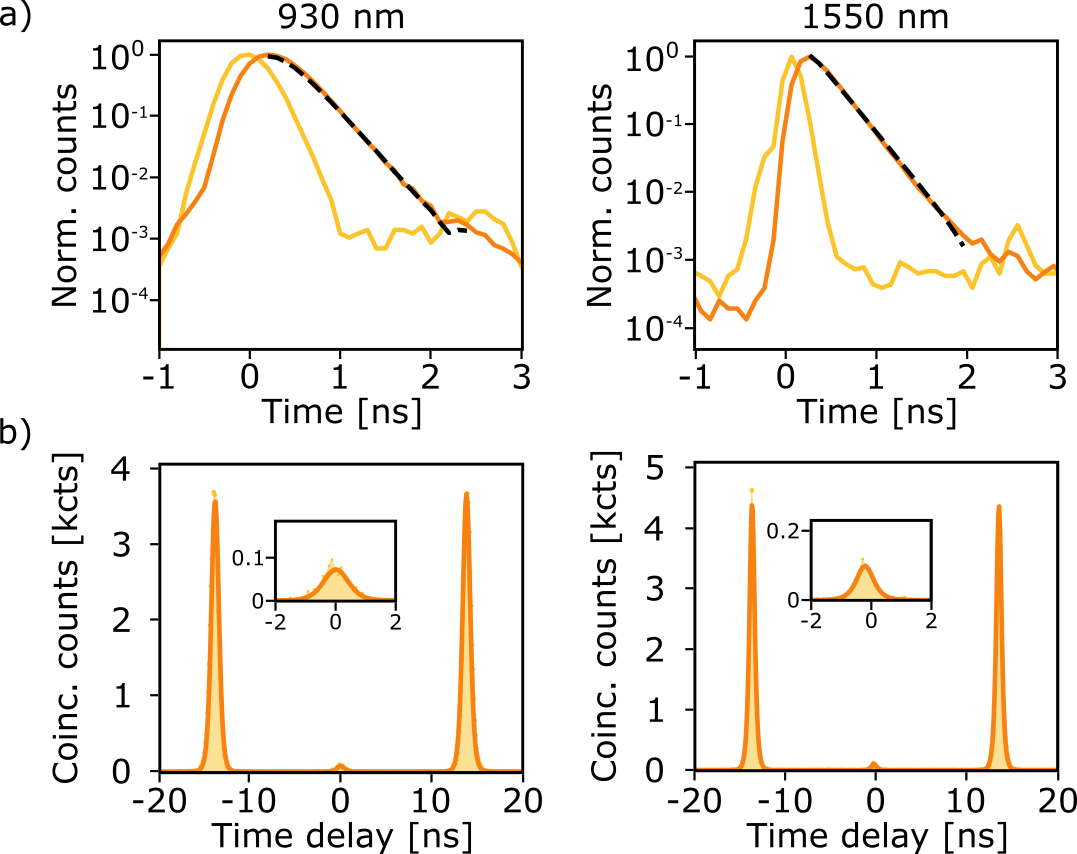}
\caption{\textbf{Measured lifetime and purity before (left) and after conversion (right). a) Lifetime measurements.} The lifetime is measured and plot in orange. The instrument response function (shown in yellow) is convolved with an exponential decay to precisely fit the single photon lifetime (fit shown in dashed black). \textbf{b) Purity measurements.} Coincidence counts are measured in a HBT setup: data is shown with yellow points while the fit is shown is orange.}
\label{fig:lifetime_g2}
\end{figure}

In a second experiment with a QD source (operated in a 4 K closed-cycle cryostat) emitting at 945 nm, the quantum characteristics of the source was investigated both before and after the the QFC process. The output wavelength was moved to 1543 nm in order to completely eliminate the noise coming from the pump laser, either direct or Raman. Fig.~\ref{fig:lifetime_g2}a) on the left shows the measured lifetime of the source (measured with the setup shown in Fig.~\ref{fig:setup}b)), the detector/instrument response function (IRF) and the fit with an exponential decay convolved with the IRF. From the fitted curve we can extract a time constant of $\tau_{930}$=(0.271 $\pm$ 0.016) ns. On the right, we report the same measurement carried out on the converted photons. We find $\tau_{1550}$=(0.269 $\pm$ 0.004) ns, confirming no change in the lifetime of the frequency converted single-photons. To assess the purity of the source, we measured the second order correlation $g^{(2)}(0)$ of the photons at 930 nm and at 1550 nm using a Hanbury Brown-Twiss (HBT) setup as shown in Fig.~\ref{fig:setup}c). The resulting coincidence counts are reported in Fig.~\ref{fig:lifetime_g2}b) in the left (right) plot before- (after-)QFC with an integration time of 10 (30) minutes. By fitting the data, we can compare the area of the central peak with the area of a far away peak (at around 40 $\mu$s delay, see supplementary materials for the long time scale behaviour). This gives us $g^{(2)}_{930}(0)=(2.0 \pm 0.3)\%$ and $g^{(2)}_{1550}(0)=(2.4 \pm 0.2)\%$ in the two wavelength regimes. The two values and error ranges show that the QFC process does not increase the multi-photon probability, i.e. the C-band single-photon source is only limited by the few percent impurity of the original QD source.

\begin{figure}
\centering
\includegraphics[width=0.45\textwidth]{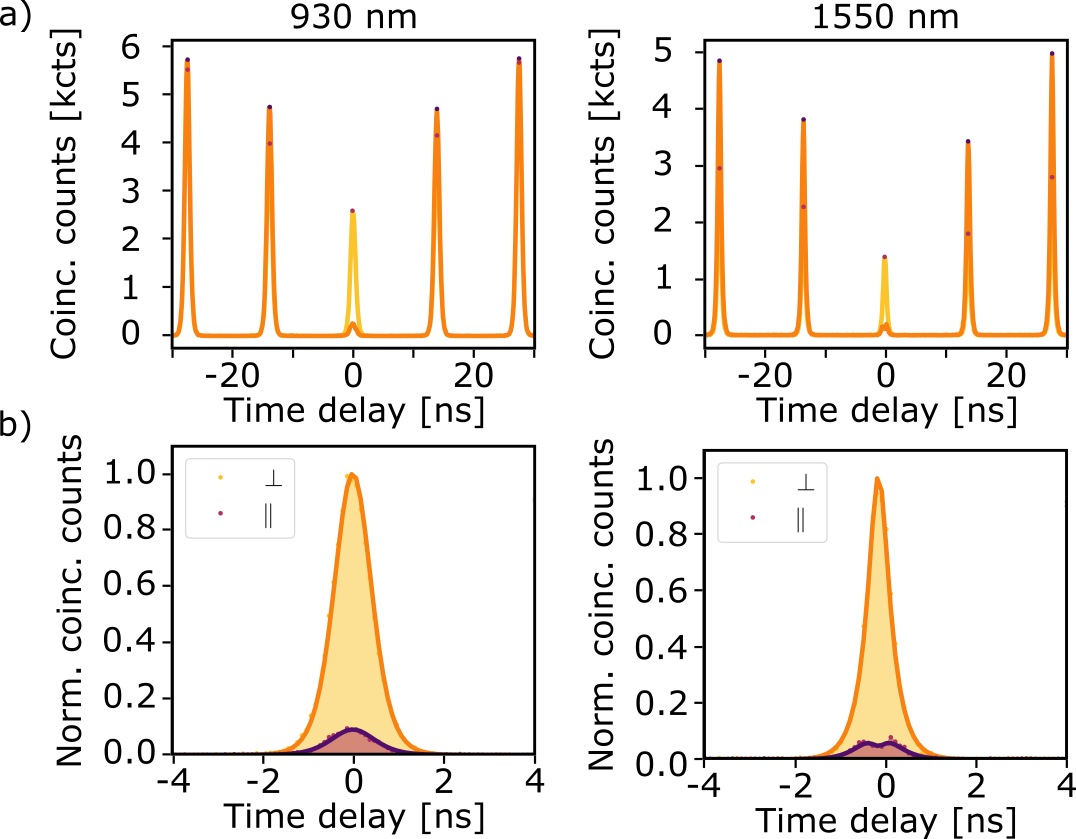}
\caption{\textbf{Measured indistinguishability before (left) and after conversion (right). a) Coincidence counts.} Data, measured in a HOM setup, is shown in orange (yellow) for the co-polarized (cross-polarized) configuration, and peak values are highlighted with blue (purple) points. \textbf{b) Zoom in view on the central peaks.} Normalized data is shown with yellow points in the cross-polarized scheme (fit shown in orange), and with purple points in the co-polarized scheme (fit shown in blue). Each trace is firstly normalized to a far away side peak area, and then both are normalized on the cross-polarization central peak for clarity.}
\label{fig:hom}
\end{figure}

Finally, we characterize the indistinguishability of the source, before and after conversion by measuring coincidence counts in a Hong-Ou Mandel (HOM) setup, shown in Fig.~\ref{fig:setup}d). Two photons generated by consecutive excitation pulses are interfered in an asymmetric Mach-Zendher interferometer and a half waveplate is used to cross-polarize the photons to make them distinguishable. The results are shown in Fig.~\ref{fig:hom}a). On the left, we report measurements from the QD source at 930 nm (with 10 minutes integration time per trace), and on the right similar measurements after conversion at 1550 nm (1 hour integration time per trace). In both cases, the raw visibility is obtained by fitting the data, normalizing each trace with the area of a peak at a delay of approximately 500 ns, and then comparing the areas of the central peaks in the co- and cross-polarized cases, $A_{\parallel}$ and $A_{\perp}$: $V^{(\text{raw})}=(A_{\perp}-A_{\parallel})/A_{\perp}$. The normalized coincidence counts are shown in Fig.~\ref{fig:hom}b) for both before and after conversion measurements for ease of comparison. The resulting raw visibilities are $V^{(\text{raw})}_{930}=(89.2 \pm 0.9)\%$ and $V^{(\text{raw})}_{1550}=(88.8 \pm 1.4)\%$. Taking into account the non zero $g^{(2)}(0)$ (assuming it is due to distinguishable photons) and the imperfections in the optical setups such as imperfect splitting ratios of the beam splitters and non-unity interference visibility (see supplementary materials), we extract the corrected visibility \cite{santori2002indistinguishable} $V^{(\text{corr})}_{930}=(93.5 \pm 1.1)\%$ and $V^{(\text{corr})}_{1550}=(94.8 \pm 1.6)\%$ before and after conversion, respectively. Both the raw and corrected visibility data reveal a high degree of indistinguishability of the QD source, which is preserved after QFC. We observe a qualitatively different shape of the suppressed HOM peak for the two sets of data, which can be attributed to the fact that detector response time at 1550 nm is significantly faster than at 930 nm, which is evident in the IRF of the 930 nm and 1550 nm detectors in Fig.~\ref{fig:lifetime_g2}a) (further analysis in the supplementary materials).

In conclusion, we have demonstrated frequency conversion of indistinguishable single photons emitted in the 900-950 nm to the telecommunication C-band around 1550 nm. We have shown simultaneously a high external efficiency of $40.8\%$ and the low-noise behaviour required to maintain the high purity and indistinguishability characteristic of the QDs used as sources, achieving $g^{(2)}(0)=2.4\%$ and $V^{\text{corr}}=94.8\%$, respectively.
The end-to-end conversion efficiency is currently limited by (apart from the optical loss of the setup components) the coupling of the converted 1550 nm photons to the output fiber, and could be readily improved by tailoring the light wavefront to achieve better mode matching, for example using spatial light modulators. Future work will address integrating QFC on-chip, using thin-film ppLN or GaAs waveguides \cite{chen2021photon,stanton2020efficient,lu2019periodically}, which are expected to further reduce the required pump power and allow to scale up the conversion to multiple emitters. The demonstrated low-noise QFC enables developing schemes for long-distance quantum communication and it offers the possibility to compensate for the inhomogeneous broadening of the QD emission wavelength, as demonstrated by several works \cite{ates2012two,weber2019two} recently achieving up to $(67 \pm 2)\%$ two-photon indistinguishability \cite{you2021quantum}.
Moreover, our schemes could be used in combination with highly-efficient photonic integrated circuits to implement device-independent quantum key distribution protocols, where maintaining photon coherence after conversion is crucial to perform Bell-state measurements at a remote heralding station \cite{kolodynski2020device,gonzalez2021violation}.
Finally, the QFC demonstrated here closes the gap between the single-photon source technology of InAs QDs in GaAs and the mature silicon-on-insulator photonic integrated circuit technology \cite{carolan2015universal,wang2018multidimensional,paesani2019generation}, enabling the modular integration of quantum communication, computing, and simulation.


\section*{Supplementary materials}
See Supplementary Materials for further description of the experimental setup, the conversion fit model and further details on the measured purity and indistinguishability of the single-photons.

\section*{Author contributions}
B.D.L., C.F., X.Z. and M.L.C. planned and carried out experimental work. X.Z.,  R.U., and H.T. designed the experimental setup. B.D.L. and C.F. analyzed the data. B.D.L. and L.M. prepared the figures and wrote the manuscript with input from all authors. S.S., A.D.W., and A.L. carried out the growth and design of the wafer. P.L. and L.M. supervised the project.

\section*{Acknowledgments}
We gratefully acknowledge financial support from the Danish National Research Foundation (Center of Excellence Hy-Q DNRF139), Innovationsfonden (No. 9090-00031B, FIRE-Q). M.L.C acknowledges support from European Union’s Horizon 2020 Research and innovation Programme under the Marie Sklodowska-Curie Grant Agreements No. 861097 (QUDOT-TECH). A.L., S.S., and A.D.W. acknowledge support of BMBF (Q.Link.X 16KIS0867) and the DFG (TRR 160).

\section*{Competing interests}
P.L. is founder and minority shareholder in the company Sparrow Quantum. The authors declare that there are no other competing interests.

\section*{Data Availability Statement}
All data needed to evaluate the conclusions in the paper are present in the paper and/or the Supplementary Materials. Additional data related to this paper available from the corresponding author upon reasonable request.

\bibliographystyle{IEEEtran}
\bibliography{Ref}

\begin{thebibliography}{10}
\providecommand{\url}[1]{#1}
\csname url@samestyle\endcsname
\providecommand{\newblock}{\relax}
\providecommand{\bibinfo}[2]{#2}
\providecommand{\BIBentrySTDinterwordspacing}{\spaceskip=0pt\relax}
\providecommand{\BIBentryALTinterwordstretchfactor}{4}
\providecommand{\BIBentryALTinterwordspacing}{\spaceskip=\fontdimen2\font plus
\BIBentryALTinterwordstretchfactor\fontdimen3\font minus
  \fontdimen4\font\relax}
\providecommand{\BIBforeignlanguage}[2]{{%
\expandafter\ifx\csname l@#1\endcsname\relax
\typeout{** WARNING: IEEEtran.bst: No hyphenation pattern has been}%
\typeout{** loaded for the language `#1'. Using the pattern for}%
\typeout{** the default language instead.}%
\else
\language=\csname l@#1\endcsname
\fi
#2}}
\providecommand{\BIBdecl}{\relax}
\BIBdecl

\bibitem{uppu2021quantum}
R.~Uppu, L.~Midolo, X.~Zhou, J.~Carolan, and P.~Lodahl, ``Quantum-dot-based
  deterministic photon--emitter interfaces for scalable photonic quantum
  technology,'' \emph{Nature Nanotechnology}, pp. 1--10, 2021.

\bibitem{uppu2020scalable}
R.~Uppu, F.~T. Pedersen, Y.~Wang, C.~T. Olesen, C.~Papon, X.~Zhou, L.~Midolo,
  S.~Scholz, A.~D. Wieck, A.~Ludwig \emph{et~al.}, ``Scalable integrated
  single-photon source,'' \emph{Science advances}, vol.~6, no.~50, p. eabc8268,
  2020.

\bibitem{appel2021coherent}
M.~H. Appel, A.~Tiranov, A.~Javadi, M.~C. L{\"o}bl, Y.~Wang, S.~Scholz, A.~D.
  Wieck, A.~Ludwig, R.~J. Warburton, and P.~Lodahl, ``Coherent spin-photon
  interface with waveguide induced cycling transitions,'' \emph{Physical Review
  Letters}, vol. 126, no.~1, p. 013602, 2021.

\bibitem{pedersen2021demand}
F.~T. Pedersen, E.~M. Gonz{\'a}lez-Ruiz, N.~Hauff, Y.~Wang, A.~D. Wieck,
  A.~Ludwig, R.~Schott, L.~Midolo, A.~S. S{\o}rensen, R.~Uppu \emph{et~al.},
  ``On-demand source of dual-rail photon pairs based on chiral interaction in a
  nanophotonic waveguide,'' \emph{arXiv preprint arXiv:2109.03519}, 2021.

\bibitem{wang2016near}
H.~Wang, Z.-C. Duan, Y.-H. Li, S.~Chen, J.-P. Li, Y.-M. He, M.-C. Chen, Y.~He,
  X.~Ding, C.-Z. Peng \emph{et~al.}, ``Near-transform-limited single photons
  from an efficient solid-state quantum emitter,'' \emph{Physical Review
  Letters}, vol. 116, no.~21, p. 213601, 2016.

\bibitem{tomm2021bright}
N.~Tomm, A.~Javadi, N.~O. Antoniadis, D.~Najer, M.~C. L{\"o}bl, A.~R. Korsch,
  R.~Schott, S.~R. Valentin, A.~D. Wieck, A.~Ludwig \emph{et~al.}, ``A bright
  and fast source of coherent single photons,'' \emph{Nature Nanotechnology},
  vol.~16, no.~4, pp. 399--403, 2021.

\bibitem{miya1979ultimate}
T.~Miya, Y.~Terunuma, T.~Hosaka, and T.~Miyashita, ``Ultimate low-loss
  single-mode fibre at 1.55 $\mu$m,'' \emph{Electronics Letters}, vol.~15,
  no.~4, pp. 106--108, 1979.

\bibitem{seravalli2007quantum}
L.~Seravalli, M.~Minelli, P.~Frigeri, S.~Franchi, G.~Guizzetti, M.~Patrini,
  T.~Ciabattoni, and M.~Geddo, ``Quantum dot strain engineering of
  {I}n{A}s/{I}n{G}a{A}s nanostructures,'' \emph{Journal of applied physics},
  vol. 101, no.~2, p. 024313, 2007.

\bibitem{anderson2020quantum}
M.~Anderson, T.~M{\"u}ller, J.~Huwer, J.~Skiba-Szymanska, A.~Krysa,
  R.~Stevenson, J.~Heffernan, D.~Ritchie, and A.~Shields, ``Quantum
  teleportation using highly coherent emission from telecom c-band quantum
  dots,'' \emph{npj Quantum Information}, vol.~6, no.~1, pp. 1--7, 2020.

\bibitem{takemoto2004non}
K.~Takemoto, Y.~Sakuma, S.~Hirose, T.~Usuki, N.~Yokoyama, T.~Miyazawa,
  M.~Takatsu, and Y.~Arakawa, ``Non-classical photon emission from a single
  {I}n{A}s/{I}n{P} quantum dot in the 1.3-$\mu$m optical-fiber band,''
  \emph{Japanese journal of applied physics}, vol.~43, no.~7B, p. L993, 2004.

\bibitem{miyazawa2005single}
T.~Miyazawa, K.~Takemoto, Y.~Sakuma, S.~Hirose, T.~Usuki, N.~Yokoyama,
  M.~Takatsu, and Y.~Arakawa, ``Single-photon generation in the 1.55-$\mu$m
  optical-fiber band from an {I}n{A}s/{I}n{P} quantum dot,'' \emph{Japanese
  Journal of Applied Physics}, vol.~44, no.~5L, p. L620, 2005.

\bibitem{ward2005demand}
M.~Ward, O.~Karimov, D.~Unitt, Z.~Yuan, P.~See, D.~Gevaux, A.~Shields,
  P.~Atkinson, and D.~Ritchie, ``On-demand single-photon source for 1.3 $\mu$m
  telecom fiber,'' \emph{Applied Physics Letters}, vol.~86, no.~20, p. 201111,
  2005.

\bibitem{takemoto2007optical}
K.~Takemoto, M.~Takatsu, S.~Hirose, N.~Yokoyama, Y.~Sakuma, T.~Usuki,
  T.~Miyazawa, and Y.~Arakawa, ``An optical horn structure for single-photon
  source using quantum dots at telecommunication wavelength,'' \emph{Journal of
  applied physics}, vol. 101, no.~8, p. 081720, 2007.

\bibitem{takemoto2008telecom}
K.~Takemoto, S.~Hirose, M.~Takatsu, N.~Yokoyama, Y.~Sakuma, T.~Usuki,
  T.~Miyazawa, and Y.~Arakawa, ``Telecom single-photon source with horn
  structure,'' \emph{physica status solidi c}, vol.~5, no.~9, pp. 2699--2703,
  2008.

\bibitem{benyoucef2013telecom}
M.~Benyoucef, M.~Yacob, J.~Reithmaier, J.~Kettler, and P.~Michler,
  ``Telecom-wavelength (1.5 $\mu$m) single-photon emission from {I}n{P}-based
  quantum dots,'' \emph{Applied Physics Letters}, vol. 103, no.~16, p. 162101,
  2013.

\bibitem{paul2015metal}
M.~Paul, J.~Kettler, K.~Zeuner, C.~Clausen, M.~Jetter, and P.~Michler,
  ``Metal-organic vapor-phase epitaxy-grown ultra-low density
  {I}n{G}a{A}s/{G}a{A}s quantum dots exhibiting cascaded single-photon emission
  at 1.3 $\mu$m,'' \emph{Applied Physics Letters}, vol. 106, no.~12, p. 122105,
  2015.

\bibitem{miyazawa2016single}
T.~Miyazawa, K.~Takemoto, Y.~Nambu, S.~Miki, T.~Yamashita, H.~Terai,
  M.~Fujiwara, M.~Sasaki, Y.~Sakuma, M.~Takatsu \emph{et~al.}, ``Single-photon
  emission at 1.5 $\mu$m from an {I}n{A}s/{I}n{P} quantum dot with highly
  suppressed multi-photon emission probabilities,'' \emph{Applied Physics
  Letters}, vol. 109, no.~13, p. 132106, 2016.

\bibitem{al2016resonance}
R.~Al-Khuzheyri, A.~C. Dada, J.~Huwer, T.~S. Santana, J.~Skiba-Szymanska,
  M.~Felle, M.~Ward, R.~Stevenson, I.~Farrer, M.~G. Tanner \emph{et~al.},
  ``Resonance fluorescence from a telecom-wavelength quantum dot,''
  \emph{Applied Physics Letters}, vol. 109, no.~16, p. 163104, 2016.

\bibitem{paul2017single}
M.~Paul, F.~Olbrich, J.~H{\"o}schele, S.~Schreier, J.~Kettler, S.~L. Portalupi,
  M.~Jetter, and P.~Michler, ``Single-photon emission at 1.55 $\mu$m from
  movpe-grown {I}n{A}s quantum dots on {I}n{G}a{A}s/{G}a{A}s metamorphic
  buffers,'' \emph{Applied Physics Letters}, vol. 111, no.~3, p. 033102, 2017.

\bibitem{muller2018quantum}
T.~M{\"u}ller, J.~Skiba-Szymanska, A.~Krysa, J.~Huwer, M.~Felle, M.~Anderson,
  R.~Stevenson, J.~Heffernan, D.~A. Ritchie, and A.~Shields, ``A quantum
  light-emitting diode for the standard telecom window around 1,550 nm,''
  \emph{Nature communications}, vol.~9, no.~1, pp. 1--6, 2018.

\bibitem{kolatschek2021bright}
S.~Kolatschek, C.~Nawrath, S.~Bauer, J.~Huang, J.~Fischer, R.~Sittig,
  M.~Jetter, S.~L. Portalupi, and P.~Michler, ``Bright purcell enhanced
  single-photon source in the telecom o-band based on a quantum dot in a
  circular bragg grating,'' \emph{Nano Letters}, vol.~21, no.~18, pp.
  7740--7745, 2021.

\bibitem{kim2016two}
J.-H. Kim, T.~Cai, C.~J. Richardson, R.~P. Leavitt, and E.~Waks, ``Two-photon
  interference from a bright single-photon source at telecom wavelengths,''
  \emph{Optica}, vol.~3, no.~6, pp. 577--584, 2016.

\bibitem{nawrath2019coherence}
C.~Nawrath, F.~Olbrich, M.~Paul, S.~Portalupi, M.~Jetter, and P.~Michler,
  ``Coherence and indistinguishability of highly pure single photons from
  non-resonantly and resonantly excited telecom c-band quantum dots,''
  \emph{Applied Physics Letters}, vol. 115, no.~2, p. 023103, 2019.

\bibitem{srocka2020deterministically}
N.~Srocka, P.~Mrowi{\'n}ski, J.~Gro{\ss}e, M.~von Helversen, T.~Heindel,
  S.~Rodt, and S.~Reitzenstein, ``Deterministically fabricated quantum dot
  single-photon source emitting indistinguishable photons in the telecom
  o-band,'' \emph{Applied Physics Letters}, vol. 116, no.~23, p. 231104, 2020.

\bibitem{intallura2007quantum}
P.~Intallura, M.~Ward, O.~Karimov, Z.~Yuan, P.~See, A.~Shields, P.~Atkinson,
  and D.~Ritchie, ``Quantum key distribution using a triggered quantum dot
  source emitting near 1.3 $\mu$m,'' \emph{Applied Physics Letters}, vol.~91,
  no.~16, p. 161103, 2007.

\bibitem{intallura2009quantum}
P.~Intallura, M.~Ward, O.~Karimov, Z.~Yuan, P.~See, P.~Atkinson, D.~Ritchie,
  and A.~Shields, ``Quantum communication using single photons from a
  semiconductor quantum dot emitting at a telecommunication wavelength,''
  \emph{Journal of Optics A: Pure and Applied Optics}, vol.~11, no.~5, p.
  054005, 2009.

\bibitem{takemoto2010transmission}
K.~Takemoto, Y.~Nambu, T.~Miyazawa, K.~Wakui, S.~Hirose, T.~Usuki, M.~Takatsu,
  N.~Yokoyama, K.~Yoshino, A.~Tomita \emph{et~al.}, ``Transmission experiment
  of quantum keys over 50 km using high-performance quantum-dot single-photon
  source at 1.5 $\mu$m wavelength,'' \emph{Applied Physics Express}, vol.~3,
  no.~9, p. 092802, 2010.

\bibitem{takemoto2015quantum}
K.~Takemoto, Y.~Nambu, T.~Miyazawa, Y.~Sakuma, T.~Yamamoto, S.~Yorozu, and
  Y.~Arakawa, ``Quantum key distribution over 120 km using ultrahigh purity
  single-photon source and superconducting single-photon detectors,''
  \emph{Scientific reports}, vol.~5, no.~1, pp. 1--7, 2015.

\bibitem{kupko2021evaluating}
T.~Kupko, L.~Rickert, F.~Urban, J.~Gro{\ss}e, N.~Srocka, S.~Rodt, A.~Musia{\l},
  K.~{\.Z}o{\l}nacz, P.~Mergo, K.~Dybka \emph{et~al.}, ``Evaluating a
  stand-alone quantum-dot single-photon source for quantum key distribution at
  telecom wavelengths,'' \emph{arXiv preprint arXiv:2105.03473}, 2021.

\bibitem{zaske2012visible}
S.~Zaske, A.~Lenhard, C.~A. Ke{\ss}ler, J.~Kettler, C.~Hepp, C.~Arend,
  R.~Albrecht, W.-M. Schulz, M.~Jetter, P.~Michler \emph{et~al.},
  ``Visible-to-telecom quantum frequency conversion of light from a single
  quantum emitter,'' \emph{Physical review letters}, vol. 109, no.~14, p.
  147404, 2012.

\bibitem{pelc2012downconversion}
J.~S. Pelc, L.~Yu, K.~De~Greve, P.~L. McMahon, C.~M. Natarajan,
  V.~Esfandyarpour, S.~Maier, C.~Schneider, M.~Kamp, S.~H{\"o}fling
  \emph{et~al.}, ``Downconversion quantum interface for a single quantum dot
  spin and 1550-nm single-photon channel,'' \emph{Optics express}, vol.~20,
  no.~25, pp. 27\,510--27\,519, 2012.

\bibitem{Kambs:16}
\BIBentryALTinterwordspacing
B.~Kambs, J.~Kettler, M.~Bock, J.~N. Becker, C.~Arend, A.~Lenhard, S.~L.
  Portalupi, M.~Jetter, P.~Michler, and C.~Becher, ``Low-noise quantum
  frequency down-conversion of indistinguishable photons,'' \emph{Opt.
  Express}, vol.~24, no.~19, pp. 22\,250--22\,260, Sep 2016. [Online].
  Available:
  \url{http://www.osapublishing.org/oe/abstract.cfm?URI=oe-24-19-22250}
\BIBentrySTDinterwordspacing

\bibitem{morrison2021bright}
C.~L. Morrison, M.~Rambach, Z.~X. Koong, F.~Graffitti, F.~Thorburn, A.~K. Kar,
  Y.~Ma, S.-I. Park, J.~D. Song, N.~G. Stoltz \emph{et~al.}, ``A bright source
  of telecom single photons based on quantum frequency conversion,''
  \emph{Applied Physics Letters}, vol. 118, no.~17, p. 174003, 2021.

\bibitem{santori2002indistinguishable}
C.~Santori, D.~Fattal, J.~Vu{\v{c}}kovi{\'c}, G.~S. Solomon, and Y.~Yamamoto,
  ``Indistinguishable photons from a single-photon device,'' \emph{nature},
  vol. 419, no. 6907, pp. 594--597, 2002.

\bibitem{chen2021photon}
J.-Y. Chen, Z.~Li, Z.~Ma, C.~Tang, H.~Fan, Y.~M. Sua, and Y.-P. Huang, ``Photon
  conversion and interaction on chip,'' \emph{arXiv preprint arXiv:2105.00275},
  2021.

\bibitem{stanton2020efficient}
E.~J. Stanton, J.~Chiles, N.~Nader, G.~Moody, N.~Volet, L.~Chang, J.~E. Bowers,
  S.~W. Nam, and R.~P. Mirin, ``Efficient second harmonic generation in
  nanophotonic {G}a{A}s-on-insulator waveguides,'' \emph{Optics express},
  vol.~28, no.~7, pp. 9521--9532, 2020.

\bibitem{lu2019periodically}
J.~Lu, J.~B. Surya, X.~Liu, A.~W. Bruch, Z.~Gong, Y.~Xu, and H.~X. Tang,
  ``Periodically poled thin-film lithium niobate microring resonators with a
  second-harmonic generation efficiency of 250,000\%/w,'' \emph{Optica},
  vol.~6, no.~12, pp. 1455--1460, 2019.

\bibitem{ates2012two}
S.~Ates, I.~Agha, A.~Gulinatti, I.~Rech, M.~T. Rakher, A.~Badolato, and
  K.~Srinivasan, ``Two-photon interference using background-free quantum
  frequency conversion of single photons emitted by an {I}n{A}s quantum dot,''
  \emph{Physical review letters}, vol. 109, no.~14, p. 147405, 2012.

\bibitem{weber2019two}
J.~H. Weber, B.~Kambs, J.~Kettler, S.~Kern, J.~Maisch, H.~Vural, M.~Jetter,
  S.~L. Portalupi, C.~Becher, and P.~Michler, ``Two-photon interference in the
  telecom c-band after frequency conversion of photons from remote quantum
  emitters,'' \emph{Nature nanotechnology}, vol.~14, no.~1, pp. 23--26, 2019.

\bibitem{you2021quantum}
X.~You, M.-Y. Zheng, S.~Chen, R.-Z. Liu, J.~Qin, M.-C. Xu, Z.-X. Ge, T.-H.
  Chung, Y.-K. Qiao, Y.-F. Jiang \emph{et~al.}, ``Quantum interference between
  independent solid-state single-photon sources separated by 300 km fiber,''
  \emph{arXiv preprint arXiv:2106.15545}, 2021.

\bibitem{kolodynski2020device}
J.~Ko{\l}ody{\'n}ski, A.~M{\'a}ttar, P.~Skrzypczyk, E.~Woodhead, D.~Cavalcanti,
  K.~Banaszek, and A.~Ac{\'\i}n, ``Device-independent quantum key distribution
  with single-photon sources,'' \emph{Quantum}, vol.~4, p. 260, 2020.

\bibitem{gonzalez2021violation}
E.~M. Gonz{\'a}lez-Ruiz, S.~K. Das, P.~Lodahl, and A.~S. S{\o}rensen,
  ``Violation of bell's inequality with quantum-dot single-photon sources,''
  \emph{arXiv preprint arXiv:2109.14712}, 2021.

\bibitem{carolan2015universal}
J.~Carolan, C.~Harrold, C.~Sparrow, E.~Mart{\'\i}n-L{\'o}pez, N.~J. Russell,
  J.~W. Silverstone, P.~J. Shadbolt, N.~Matsuda, M.~Oguma, M.~Itoh
  \emph{et~al.}, ``Universal linear optics,'' \emph{Science}, vol. 349, no.
  6249, pp. 711--716, 2015.

\bibitem{wang2018multidimensional}
J.~Wang, S.~Paesani, Y.~Ding, R.~Santagati, P.~Skrzypczyk, A.~Salavrakos,
  J.~Tura, R.~Augusiak, L.~Man{\v{c}}inska, D.~Bacco \emph{et~al.},
  ``Multidimensional quantum entanglement with large-scale integrated optics,''
  \emph{Science}, vol. 360, no. 6386, pp. 285--291, 2018.

\bibitem{paesani2019generation}
S.~Paesani, Y.~Ding, R.~Santagati, L.~Chakhmakhchyan, C.~Vigliar, K.~Rottwitt,
  L.~K. Oxenl{\o}we, J.~Wang, M.~G. Thompson, and A.~Laing, ``Generation and
  sampling of quantum states of light in a silicon chip,'' \emph{Nature
  Physics}, vol.~15, no.~9, pp. 925--929, 2019.

\end{thebibliography}

\end{document}